\documentclass[]{aa}

\usepackage{graphicx}
\usepackage{txfonts}
\usepackage{lipsum}
\usepackage{subcaption}
\usepackage{lscape}
\usepackage{placeins}
\usepackage[colorlinks=true,linkcolor=blue,citecolor=blue]{hyperref}
\usepackage{threeparttablex}
\usepackage{multirow}

\begin{document}

   \title{Validating the bright Gaia celestial reference frame with new VLBI astrometry of radio stars}
   \titlerunning{Validating bright Gaia-CRF with VLBI radio star astrometry}

   \author{J. Zhang\inst{1,2},
           B. Zhang\inst{1}\fnmsep\thanks{Corresponding author; \email{zb@shao.ac.cn}},
           S. Xu\inst{3},
           \and X. Mai\inst{1,2,4}
          }
    \authorrunning{J. Zhang et al.}

   \institute{Shanghai Astronomical Observatory, Chinese Academy of Sciences, 80 Nandan Road, Shanghai 200030, People’s Republic of China
              \and University of Chinese Academy of Sciences, No.19 (A) Yuquan Rd, Shijingshan, Beijing 100049, People’s Republic of China
              \and Korea Astronomy and Space Science Institute, 776 Daedeok-daero, Yuseong-gu, Daejeon 34055, Republic of Korea
              \and University of Helsinki, P.O. Box 64, FI-00014, Finland
             }
 
  \abstract
   {}
   {There exist inconsistencies between the bright and faint Gaia Celestial Reference Frame 3 (Gaia-CRF3), which manifests as a systematic rotation and needs to be independently estimated then corrected in future data releases.}
   {We collected 64 radio stars with very long baseline interferometry (VLBI) astrometry, of which 16 have new VLBI observations with reference epochs close to Gaia.
   We estimated the orientation and spin biases of the bright Gaia-CRF3 by comparing VLBI radio star astrometry with their Gaia DR3 counterparts.
   We also attempted to estimate orientation by utilizing the a priori magnitude-dependent spin parameters derived from Gaia's internal estimations.}
   {
   Our independent estimation of the orientation at $G<10.5$ is $[-15\pm119, +330\pm139,+218\pm109]\,\mathrm{\mu as}$ (J2016.0), and the spin ($[+21\pm18, +52\pm20,-7\pm20]\,\mathrm{\mu as\,yr^{-1}}$) agrees with Gaia's internal estimation within a 1$\sigma$ range.
   The orientation-only estimation suggests that the orientation bias of the bright Gaia-CRF3 may also be magnitude-dependent.
   }
   {}

   \keywords{astrometry --
             reference systems --
             proper motions
            }

   \maketitle

\section{Introduction}
\label{sect:intro}
The Gaia mission \citep{2016A&A...595A...1G} provides precise astrometry for 1.8 billion sources in its latest data release, Gaia DR3 \citep{2023A&A...674A...1G}, and established its celestial reference frame, Gaia-CRF3 \citep{2022A&A...667A.148G}, based on an extragalactic source (quasar) sample of 1.6 million.
However, most quasars are faint in the optical band, the $G$-band magnitude of which ranges from 13 to 21, without coverage of the bright end ($G<13$). This may result in potential inconsistency at the bright end.
Unfortunately, evidence suggests that this inconsistency exists and manifests as a systematic rotation (orientation bias at reference epoch and spin with time) \citep{2018A&A...616A...2L,2018ApJS..239...31B}.
The origin of this rotation may be astrometric instrument calibration limitations, such as window class effects \citep{2020A&A...633A...1L}.

The rotation can be estimated internally within Gaia or through a comparison with the Hipparcos catalog \citep{2007ASSL..350.....V}.
For example, \citet{2021A&A...649A.124C} estimated spin parameters with samples collected from known binaries and open clusters, and an ad hoc correction is applied in the astrometric solution of Gaia DR3 based on a comparison of proper motions with Hipparcos \citep{2021A&A...649A...2L}.

It is necessary to validate the above estimations with independent external methods.
For example, very long baseline interferometry (VLBI) astrometry of radio stars is a direct approach for the validation \citep{2016MNRAS.461.1937M}.
Gaia-CRF3 is aligned to the International Celestial Reference Frame 3 \citep[ICRF3,][]{2020A&A...644A.159C} using common quasars at the faint end.
Although there are still inconsistencies in a small number of common quasars \citep{2025ivs..conf..300C}, the overall alignment uncertainty is about 7\,$\mu$as \citep{2022A&A...667A.148G}.
Since VLBI astrometry of radio stars is on an ICRF base, a comparison between VLBI and Gaia astrometry of radio stars can give an estimation of orientation and spin parameters between the bright and faint end of Gaia-CRF.
There have been several related works, such as \citet{2020A&A...633A...1L}, \citet{2022AstL...48..790B}, \citet{2023A&A...676A..11L}, \citet{2024A&A...689A.134L}, and \citet{2024MNRAS.529.2062Z}, but the small sample size of available radio stars limits the reliability of their results.

In our previous studies, we observed 16 radio stars in total with the Very Long Baseline Array (VLBA), which significantly expanded the available radio star sample \citep[Jiang et al. in prep., Zhang et al. in prep.]{2023MNRAS.524.5357C}.
In this paper, we present a new CRF alignment result based on radio stars, integrating historical data with our new data.
The data we use are introduced in Sect. \ref{sect:data}. Sect. \ref{sect:align} describes the methods and results of the CRF alignment. We sum up in Sect. \ref{sect:summary}.

\section{Data}
\label{sect:data}

\subsection{VLBI data}
\label{sect:vlbi}

The VLBI data used in this study consist of two parts: historical data and our new data.
\citet{2020A&A...633A...1L} collected 41 radio stars that have both VLBI parallaxes and proper motions and full astrometry in Gaia DR2.
However, some of the data are old and the positions of their calibrators are out of date.
\citet{2023A&A...676A..11L} transferred these positions to the latest ICRF3, and provided new one-epoch VLBA observations of 32 radio stars (in which one is fainter than $G=13$, and therefore not used in this study).
We included all these data in our dataset for this study.
In \citet{2023MNRAS.524.5357C}, Jiang et al. in prep., and Zhang et al. in prep., we provide VLBI astrometry of 2, 3, and 11 radio stars, respectively.
These data include 15 stars with full five-parameter astrometry and 1 with two one-epoch observations, and are all included in our dataset.
All historical and new VLBI data are listed in Appendix \ref{app:data}, including 64 radio stars.
Note that some stars have multiple data sources (e.g., we reobserved several ``old'' stars).

\subsection{Gaia data}
\label{sect:gaia}

We collected Gaia counterparts of our VLBI samples, all of which have full five-parameter astrometry in Gaia DR3, and their $G$ magnitudes range from 4.5 to 13.
The parallax zero-point problem of Gaia DR3 may affect the usage of one-epoch VLBI data, so we corrected it using the recipe provided by \citet{2021A&A...649A...4L}, with coefficients given by \citet{2024AA...691A..81D}.

\section{Estimation of the rotation between the bright Gaia-CRF3 and ICRF3}
\label{sect:align}

\subsection{Solution of both orientation and spin parameters}
\label{sect:solution_all}

We applied the CRF alignment solution method introduced in Sect. 2 of \citet{2024MNRAS.529.2062Z} to the radio star dataset.
In simple terms, we assumed that there exists a time-dependent rotation between the bright Gaia-CRF3 and ICRF3, and that it can be described with a set of six parameters $\boldsymbol{x}=[\varepsilon_{X}(T), \varepsilon_{Y}(T), \varepsilon_{Z}(T), \omega_{X}, \omega_{Y}, \omega_{Z}]^{\prime}$, of which the first three parameters describe an orientation bias at reference epoch $T=J2016.0$, while the last three parameters describe a spin at a constant angular velocity (or: the time derivative of orientation).
This rotation can be estimated from the differences between VLBI and Gaia astrometry of the same sources.
Because the rotation is on a small scale (hundreds of microarcseconds for orientation and tens of microarcseconds per year for spin), a linear approximation satisfies the need for accuracy, and we applied the least-square method to solve the parameters.

\begin{figure*}[ht]
    \centering
    \includegraphics[width=0.97\hsize]{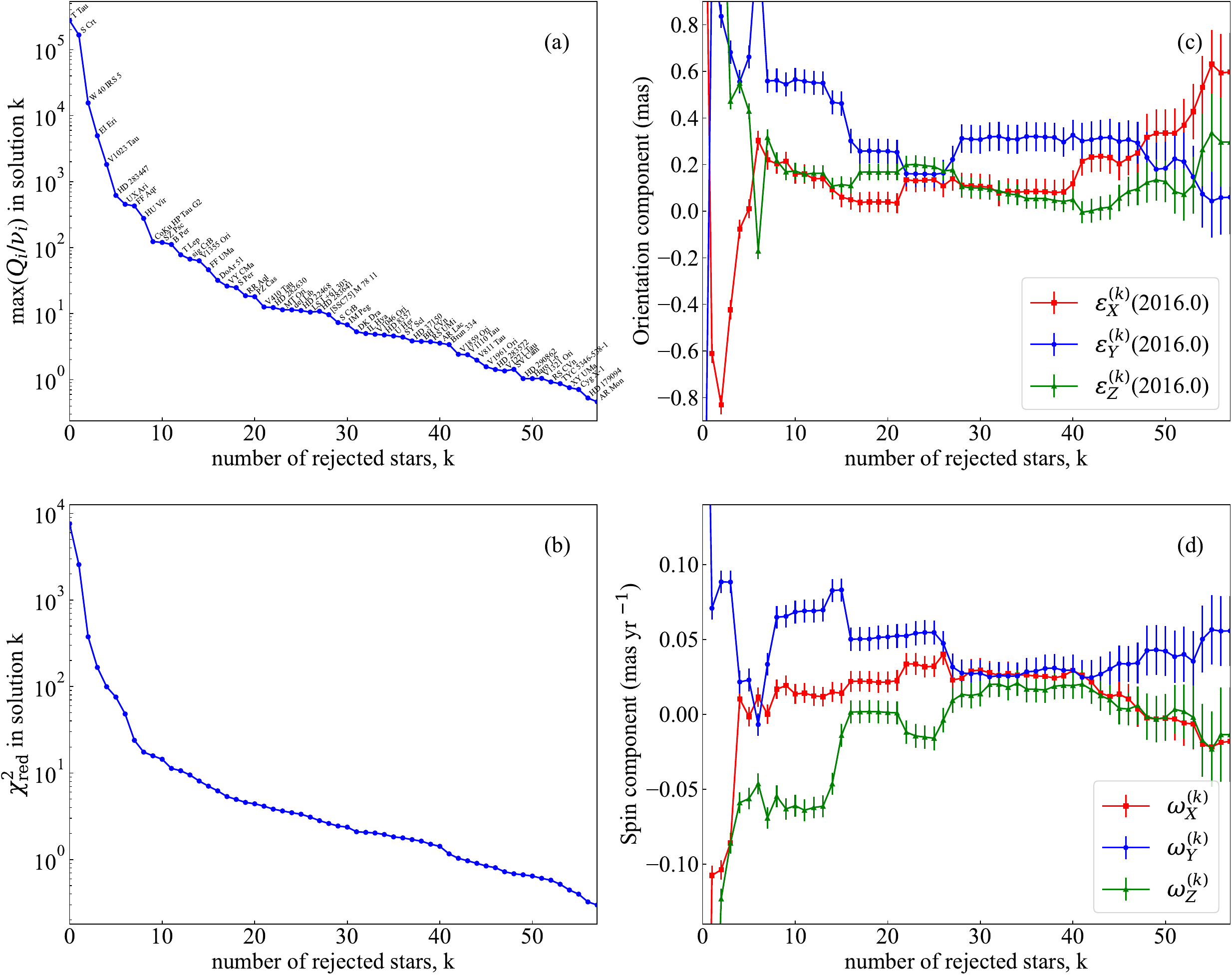}
    \caption{Evolution of loss and rotation parameters with rejected star number, $k$, increasing in the full dataset ($0<G<13$).
    (a) Max $Q_{i}(\boldsymbol{x})$ in the solution (star names labeled).
    (b) $\chi^2_{\mathrm{red}}$ of the solution.
    (c) Estimated orientation parameters.
    (d) Estimated spin parameters.
    The error bars are formal uncertainties directly derived from the covariance matrix.}
    \label{fig:evo_all}
\end{figure*}

Not all radio stars are suitable for the CRF alignment solution, for different reasons, such as the intrinsic physical properties of the stars.
For these stars, there may exist quite large differences between VLBI and Gaia astrometry, which will interfere with the estimation of CRF rotation.
To eliminate these radio stars with high loss (outliers), we applied a similar iteration procedure to the one \citet{2020A&A...633A...1L}, \citet{2023A&A...676A..11L}, \citet{2024MNRAS.529.2062Z}, and \citet{2024A&A...689A.134L} used.
The solution was first applied to the entire dataset ($0<G<13$) including $m=64$ stars, then the star $i\ (i=1\cdots m)$ with the highest loss ($Q_{i}/\nu_i$, normalized by the dimension of VLBI data) was rejected and the solution was applied again to the remaining stars.
The iteration was repeated for $k=0, 1, \cdots$ rejected stars. The evolutions of $\mathrm{max}(Q_{i}/\nu_i)$, reduced chi-square ($\chi^2_{\mathrm{red}}$), and estimated rotation parameters are shown in Figure \ref{fig:evo_all}.

In the initial rounds of iteration, due to the influence of outliers, $\chi^2_{\mathrm{red}}$ is very large and the rotation parameters also change drastically.
As $k$ increases, $\chi^2_{\mathrm{red}}$ and rotation parameters decrease and tend to be stable, and the max $Q_{i}/\nu_i$ drops below 30 when $k=19$, at which time the maximum in the remaining samples is 26.18 (VY CMa).
The rotation parameters also keep stable between $k=17$ and 21, so we adopted the result at $k=17$ for the solution, with a remaining accepted star number of $l=47$ and $\chi^2_{\mathrm{red}}=5.346$, as is shown in Table \ref{table:result}.
Compared to the results of \citet{2024A&A...689A.134L}, the uncertainties of the parameters are improved by roughly $50\%$ and $30\%$ in orientation and spin, respectively, benefitting from the additional VLBI data.

However, systematic inconsistencies do not only exist between magnitudes above and below $G=13$ in Gaia DR3.
As is shown in Fig. 4 of \citet{2021A&A...649A.124C} and Fig. 20 of \citet{2021A&A...649A...4L}, both proper motion and parallax biases are not stable between $G=10.5$ and 13.
The magnitude dependency may explain why the result of solution $0<G<13$ does not agree very well with that of \citet{2024A&A...689A.134L}, especially in orientation parameters.
The new astrometry we added changes the magnitude distribution of the radio star sample, and due to the epochs being close to the Gaia reference epoch, the new data have a high weight in the orientation estimation, which has a great impact on the result.

Internal estimations of the relationships between CRF spin and $G$ magnitude can be conducted by dividing binary and open cluster samples into multiple magnitude bins, but the number of VLBI radio star samples is too small to do so.
Therefore, we tried to solve rotation parameters with a magnitude limit of $G<10.5$, and the remaining star number is 50.
A similar iteration was conducted (iteration shown in Appendix \ref{app:iter_10.5}, statistics shown in Appendix \ref{app:stat_10.5}), and we adopted the result at $k=13$ for the solution, with a remaining star number of $l=37$ and $\chi^2_{\mathrm{red}}=5.627$.

The results of solution $0<G<10.5$ are also listed in Table \ref{table:result} and agree well with \citet{2021A&A...649A.124C}: the spin parameters are within a 1$\sigma$ range of the internal estimation at $G<10.5$.
The agreement between internal and external estimations is quite satisfying and is strong proof of the existence of systematic rotation between the bright and faint (quasar-based) Gaia-CRF3 (which is precisely aligned with ICRF3).

\subsection{Solution of orientation parameters only}
\label{sect:solution_ori}

We noticed that the discrepancies in orientation parameters between solution $0<G<13$ and $0<G<10.5$ are much beyond 1$\sigma$, and so we conducted further investigations into it.
It is necessary to fix spin parameters, then we can study orientation parameters independently.
Since the spin parameters of solution $G<10.5$ agree well with \citet{2021A&A...649A.124C}, we tend to accept that their results are reliable.
We therefore tried to solve orientation parameters with spin parameters from \citet[Table 1]{2021A&A...649A.124C} as a priori values.
This approach has several advantages:
\begin{enumerate}
    \item A priori values are given as a set of spin parameters for each of the 12 magnitude bins, which better reflects reality than using a set of magnitude-independent spin parameters;
    \item Only three orientation parameters need to be solved in the solution, reducing the negative impact of the limited sample size;
    \item Similar to (2), now we can try to solve for several magnitude bins to examine the relationship between CRF orientation and the $G$ magnitude.
\end{enumerate}

\begin{figure}[ht]
    \centering
    \includegraphics[width=0.97\hsize]{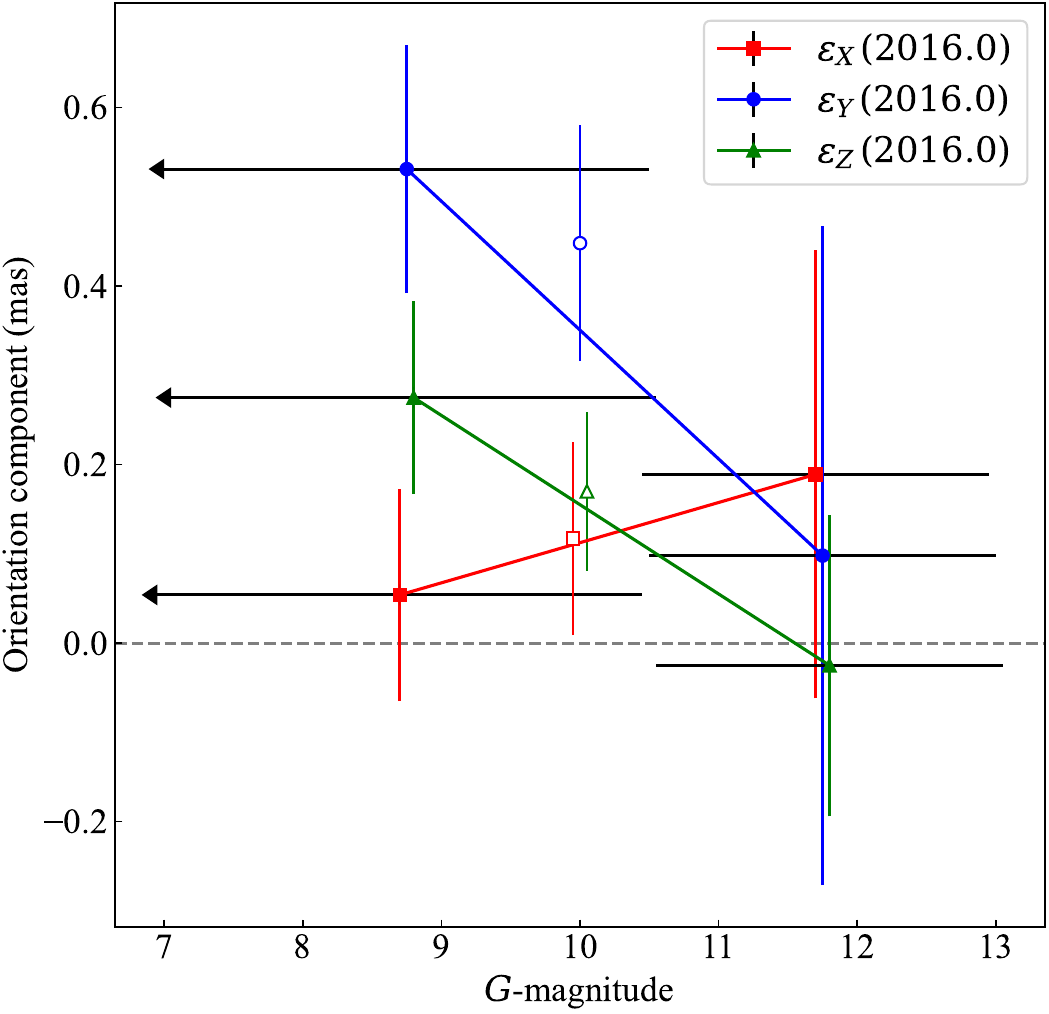}
    \caption{Orientation-magnitude relationship of orientation-only solutions.
    Hollow markers denote the solution $0<G<13$.
    Markers are horizontally shifted by 0.05 to avoid error bars overlapping.
    Horizontal lines of markers represent their magnitude range.}
    \label{fig:ori_gmag}
\end{figure}

The mathematical form of the solution needs to be slightly modified; this is introduced in Appendix \ref{app:math}.
We solved for $0<G<13$, $0<G<10.5$, and $10.5<G<13$, respectively.
The sky distribution of the samples is shown in Appendix \ref{app:distri}, and the results are listed in Table \ref{table:result}.
As is shown in Fig. \ref{fig:ori_gmag}, the results show that the orientation parameters change with $G$ magnitude, which suggests that the orientation bias of the bright Gaia-CRF3 may also be magnitude-dependent.
However, due to the limited sample size of each bin (especially $10.5<G<13$), the uncertainties are relatively large.
The results may also be biased by systematic offsets of some samples; for example, radio-optical offsets of the stars or position errors of calibrators.
Therefore, additional VLBI astrometry of radio stars is essential to achieve more robust and conclusive results.

\begin{table*}[ht]
    \caption{\label{table:result}Results of CRF alignment solution}
    \centering
    \setlength{\tabcolsep}{2pt}
    \begin{tabular}{crrrrrrrrc}
        \hline\hline
        Solution & \makebox[10pt][c]{$k$} & \makebox[10pt][c]{$l$} & \multicolumn{3}{c}{Orientation at $T={\rm J}2016.0$ (mas)} & \multicolumn{3}{c}{Spin (mas yr$^{-1}$)} & $\chi^2_{\mathrm{red}}$ \\
        & & & \makebox[52pt][c]{$\varepsilon_X(T)$} & \makebox[52pt][c]{$\varepsilon_Y(T)$} & \makebox[52pt][c]{$\varepsilon_Z(T)$} & \makebox[52pt][c]{$\omega_X$} & \makebox[52pt][c]{$\omega_Y$} & \makebox[52pt][c]{$\omega_Z$} & \\
        \hline
        $0<G<13$ & 17 & 47 & $+$0.038$\pm$0.043 & $+$0.257$\pm$0.053 & $+$0.167$\pm$0.036 & $+$0.022$\pm$0.007 & $+$0.050$\pm$0.008 & $+$0.002$\pm$0.008 & 5.346 \\
        & & & $\pm$0.099 & $\pm$0.122 & $\pm$0.082 & $\pm$0.016 & $\pm$0.018 & $\pm$0.018 & \\
        $0<G<10.5$ & 13 & 37 & $-$0.015$\pm$0.050 & $+$0.330$\pm$0.059 & $+$0.218$\pm$0.046 & $+$0.021$\pm$0.007 & $+$0.052$\pm$0.008 & $-$0.007$\pm$0.009 & 5.627 \\
        & & & $\pm$0.119 & $\pm$0.139 & $\pm$0.109 & $\pm$0.018 & $\pm$0.020 & $\pm$0.020 & \\
        \hline
        $0<G<9$ & & & \multicolumn{3}{c}{\multirow{4}*{spin-only, \citet{2021A&A...649A.124C}}} & $+$0.018$\pm$0.005 & $+$0.034$\pm$0.005 & $-$0.011$\pm$0.006 & \\
        $9<G<9.5$ & & & & & & $+$0.014$\pm$0.005 & $+$0.031$\pm$0.006 & $-$0.019$\pm$0.006 & \\
        $9.5<G<10$ & & & & & & $+$0.013$\pm$0.005 & $+$0.031$\pm$0.005 & $-$0.012$\pm$0.005 & \\
        $10<G<10.5$ & & & & & & $+$0.014$\pm$0.004 & $+$0.036$\pm$0.004 & $-$0.011$\pm$0.004 & \\
        \hline
        $0<G<13$ & 16 & 45 & $+$0.117$\pm$0.041 & $+$0.448$\pm$0.051 & $+$0.170$\pm$0.034 & & & & 6.718 \\
        & & & $\pm$0.108 & $\pm$0.132 & $\pm$0.089 & & & & \\
        $0<G<10.5$ & 13 & 34 & $+$0.054$\pm$0.048 & $+$0.531$\pm$0.056 & $+$0.275$\pm$0.043 & \multicolumn{3}{c}{orientation-only solution} & 6.221 \\
        & & & $\pm$0.119 & $\pm$0.139 & $\pm$0.108 & \multicolumn{3}{c}{adopting spin from \citet{2021A&A...649A.124C}} & \\
        $10.5<G<13$ & 3 & 11 & $+$0.189$\pm$0.089 & $+$0.098$\pm$0.131 & $-$0.025$\pm$0.060 & & & & 7.914 \\
        & & & $\pm$0.251 & $\pm$0.369 & $\pm$0.169 & & & & \\
        \hline
    \end{tabular}
    \tablefoot{The uncertainties in the second row of each solution are multiplied by $\sqrt{\chi^2_{\mathrm{red}}}$.
    Correlation coefficients are listed in Appendix \ref{app:corr}.
    }
\end{table*}

\section{Summary}
\label{sect:summary}

We have given an external estimation of the orientation and spin biases of the bright Gaia-CRF3 by comparing VLBI radio star astrometry and their Gaia DR3 counterparts.
We collected all available VLBI astrometry for 64 radio stars, including historical and new observations, and found:
\begin{enumerate}
    \item Our external estimation of spin parameters at $G<10.5$ ($[+21\pm18, +52\pm20,-7\pm20]\,\mathrm{\mu as\,yr^{-1}}$) agrees with Gaia's internal estimation \citep[binaries and open clusters,][]{2021A&A...649A.124C} within a 1$\sigma$ range;
    \item Taking spin parameters from the internal estimation as a priori values, our estimation of the orientation parameters suggests that the orientation bias of the bright Gaia-CRF3 may also be magnitude-dependent.
\end{enumerate}

External estimation with radio stars is now the only way to determine the orientation of the bright Gaia-CRF.
Our analysis demonstrates that the orientation in the bright-end Gaia-CRF samples exhibits a possible magnitude dependency.
While the present sample size of 11 stars in the range of $10.5 < G < 13$ provides a foundational insight, it underscores the need for additional radio stars with VLBI astrometry.
Drawing parallels from the Gaia parallax zero-point, which is known to vary with $G$ magnitude, color, sky position, and other potential factors, we emphasize the importance of a more even sky distribution and covering different magnitudes uniformly in future observations.
Such efforts are expected to not only refine uncertainties in parameter estimation but also enable optimized binning strategies and mitigate potential selection biases.

\begin{acknowledgements}

The Python code for CRF link is available at \citet{jingdong_zhang_2025_15031704}.

This work is supported by the National Natural Science Foundation of China (NSFC) under grant Ns. U2031212, the National Key R\&D Program of China (No. 2024YFA1611500), and the Strategic Priority Research Program of the Chinese Academy of Sciences, Grant No. XDA0350205.

This work has made use of data from the European Space Agency (ESA) mission
Gaia (\url{https://www.cosmos.esa.int/gaia}), processed by the Gaia
Data Processing and Analysis Consortium (DPAC,
\url{https://www.cosmos.esa.int/web/gaia/dpac/consortium}). Funding for the DPAC
has been provided by national institutions, in particular the institutions
participating in the Gaia Multilateral Agreement.
The Gaia services (\url{https://gaia.ari.uni-heidelberg.de/index.html}) provided by the Astronomisches Rechen-Institut (ARI) of the University of Heidelberg are used in Gaia data retrieval.
The Python package for Gaia DR3 parallax zero-point correction developed by P. Ramos can be found at \url{https://gitlab.com/icc-ub/public/gaiadr3_zeropoint}, and the coefficients used in this paper can be found at \url{https://github.com/yedings/Parallax-bias-correction-in-the-Galactic-plane}.

This work has made use of the SIMBAD database, operated at CDS, Strasbourg, France \citep[\url{https://simbad.u-strasbg.fr/simbad/}]{simbad}.

Python packages used in this work (in alphabetical order):
Astropy \citep{astropy2022}, Matplotlib \citep{2007CSE.....9...90H}, Numpy \citep{harris2020array}, and Pandas \citep{2022zndo...3509134T}.

\end{acknowledgements}

\bibliographystyle{aa}
\bibliography{main}

\begin{appendix}

\FloatBarrier
\onecolumn
\begin{landscape}
\section{VLBI data used for CRF alignment}
\label{app:data}

\begin{ThreePartTable}
\begin{longtable}{lcrrrrrrrr}
    \caption{Five-parameter VLBI data used for CRF alignment}
    \label{table:data5}\\
    \hline\hline
    \makebox[62pt][c]{Star} & Epoch & \makebox[73pt][c]{RA} & \makebox[72pt][c]{DEC} & \makebox[26pt][c]{$\sigma_{\alpha*}$} & \makebox[26pt][c]{$\sigma_{\delta}$} & \makebox[55pt][c]{$\varpi$} & \makebox[55pt][c]{$\mu_{\alpha*}$} & \makebox[55pt][c]{$\mu_\delta$} & Ref. \\
     & & \makebox[73pt][c]{(h:m:s)} & \makebox[72pt][c]{(d:m:s)} & \makebox[26pt][c]{(mas)} & \makebox[26pt][c]{(mas)} & \makebox[55pt][c]{(mas)} & \makebox[55pt][c]{(mas yr$^{-1}$)} & \makebox[55pt][c]{(mas yr$^{-1}$)} & \\
    \hline
    \endfirsthead
    \caption{continued.}\\
    \hline\hline
    \makebox[62pt][c]{Star} & Epoch & \makebox[73pt][c]{RA} & \makebox[72pt][c]{DEC} & \makebox[26pt][c]{$\sigma_{\alpha*}$} & \makebox[26pt][c]{$\sigma_{\delta}$} & \makebox[55pt][c]{$\varpi$} & \makebox[55pt][c]{$\mu_{\alpha*}$} & \makebox[55pt][c]{$\mu_\delta$} & Ref. \\
     & & \makebox[73pt][c]{(h:m:s)} & \makebox[72pt][c]{(d:m:s)} & \makebox[26pt][c]{(mas)} & \makebox[26pt][c]{(mas)} & \makebox[55pt][c]{(mas)} & \makebox[55pt][c]{(mas yr$^{-1}$)} & \makebox[55pt][c]{(mas yr$^{-1}$)} & \\
    \hline
    \endhead
    \hline
    \endfoot
    SY Scl & 2006.828200 & 00:07:36.2468069 & $-$25:29:40.028423 & 15.138 & 7.736 & 0.750$\pm$0.030 & 5.570$\pm$0.040 & $-$7.320$\pm$0.120 & 1 \\
    HD 8357 & 2023.250000 & 01:22:56.9041707 & $+$07:25:14.704379 & 0.351 & 0.738 & 22.638$\pm$0.366 & 93.885$\pm$0.223 & 230.683$\pm$0.483 & 2 \\
    S Per & 2000.888400 & 02:22:51.7103447 & $+$58:35:11.443414 & 8.678 & 9.040 & 0.413$\pm$0.017 & -0.490$\pm$0.230 & -1.190$\pm$0.200 & 3 \\
    LS I +61 303 & 1992.000000 & 02:40:31.6644017 & $+$61:13:45.596676 & 0.291 & 0.573 & 0.260$\pm$0.610 & 0.967$\pm$0.260 & -1.210$\pm$0.320 & 4 \\
    RZ Cas & 2021.125966 & 02:48:55.5212829 & $+$69:38:04.232909 & 0.238 & 0.234 & 15.274$\pm$0.093 & 2.965$\pm$0.318 & 37.543$\pm$0.185 & 5 \\
    UX Ari & 1991.247100 & 03:26:35.3613614 & $+$28:42:55.225562 & 0.444 & 0.495 & 19.370$\pm$0.390 & 41.617$\pm$0.186 & -104.010$\pm$0.200 & 4 \\
    UX Ari & 1999.986300 & 03:26:35.3838431 & $+$28:42:54.275756 & 1.313 & 0.918 & 19.900$\pm$0.500 & 44.960$\pm$0.130 & -102.330$\pm$0.090 & 6 \\
    HD 22468 & 1992.000000 & 03:36:47.3072102 & $+$00:35:17.236072 & 0.406 & 0.401 & 33.880$\pm$0.470 & -31.588$\pm$0.330 & -161.690$\pm$0.310 & 4 \\
    V1271 Tau & 2013.000000 & 03:43:48.3513936 & $+$25:00:15.202980 & 0.110 & 0.171 & 7.418$\pm$0.025 & 19.860$\pm$0.050 & -45.410$\pm$0.160 & 7 \\
    V811 Tau & 2013.000000 & 03:45:21.2049588 & $+$23:43:38.308879 & 0.118 & 0.198 & 7.223$\pm$0.057 & 19.470$\pm$0.110 & -44.390$\pm$0.270 & 7 \\
    EI Eri & 2023.250000 & 04:09:40.9526282 & $-$07:53:31.731358 & 0.229 & 0.374 & 17.894$\pm$0.146 & 36.433$\pm$0.097 & 111.925$\pm$0.109 & 2 \\
    HD 283447 & 1993.880900 & 04:14:12.9195008 & $+$28:12:12.439088 & 0.585 & 0.895 & 6.740$\pm$0.250 & 0.423$\pm$0.291 & -23.250$\pm$0.280 & 4 \\
    HD 283447 & 2006.970000 & 04:14:12.9224442 & $+$28:12:12.170880 & 0.405 & 0.522 & 7.700$\pm$0.190 & 17.092$\pm$0.077 & -24.030$\pm$0.053 & 8 \\
    V410 Tau & 2015.760000 & 04:18:31.1187088 & $+$28:27:15.760309 & 0.062 & 0.078 & 7.751$\pm$0.027 & 8.703$\pm$0.017 & -24.985$\pm$0.020 & 9 \\
    V1023 Tau & 2005.352500 & 04:18:47.0324242 & $+$28:20:07.379058 & 0.054 & 0.244 & 7.530$\pm$0.030 & 4.300$\pm$0.050 & -28.900$\pm$0.300 & 10 \\
    HD 283572 & 2005.352500 & 04:21:58.8520402 & $+$28:18:06.371139 & 0.305 & 0.094 & 7.780$\pm$0.040 & 8.880$\pm$0.060 & -26.600$\pm$0.100 & 10 \\
    HD 283572 & 2005.360000 & 04:21:58.8520428 & $+$28:18:06.370962 & 0.073 & 0.079 & 7.841$\pm$0.057 & 9.023$\pm$0.061 & -26.445$\pm$0.077 & 9 \\
    T Tau & 2004.623100 & 04:21:59.4250676 & $+$19:32:05.715231 & 0.145 & 0.633 & 6.820$\pm$0.030 & 4.020$\pm$0.030 & -1.180$\pm$0.050 & 11 \\
    HD 283641 & 2015.950000 & 04:24:49.0581756 & $+$26:43:10.128182 & 0.106 & 0.129 & 6.285$\pm$0.070 & 10.913$\pm$0.037 & -16.772$\pm$0.044 & 9 \\
    V1110 Tau & 2015.570000 & 04:34:39.2195622 & $+$25:01:00.842376 & 0.230 & 0.322 & 11.881$\pm$0.149 & -52.705$\pm$0.062 & -11.321$\pm$0.066 & 9 \\
    HD 282630 & 2016.580000 & 04:55:36.9753264 & $+$30:17:54.700203 & 0.188 & 0.228 & 7.061$\pm$0.125 & 3.897$\pm$0.113 & -24.210$\pm$0.132 & 9 \\
    T Lep & 2005.642700 & 05:04:50.8401301 & $-$21:54:16.485482 & 30.255 & 13.266 & 3.060$\pm$0.040 & 14.600$\pm$0.500 & -35.430$\pm$0.790 & 12 \\
    V1859 Ori & 2023.250000 & 05:22:54.7948822 & $+$08:58:04.463601 & 0.208 & 0.270 & 3.545$\pm$0.085 & 1.242$\pm$0.056 & -9.372$\pm$0.034 & 2 \\
    V1961 Ori & 2014.860000 & 05:34:27.3387259 & $-$05:24:22.299151 & 0.141 & 0.255 & 2.533$\pm$0.027 & -7.220$\pm$0.060 & -0.990$\pm$0.080 & 13 \\
    Brun 334 & 2015.180000 & 05:34:39.7600265 & $-$05:24:25.604141 & 0.138 & 0.275 & 2.591$\pm$0.046 & -4.010$\pm$0.080 & -1.170$\pm$0.070 & 13 \\
    V1321 Ori & 2015.180000 & 05:35:04.3011976 & $-$05:08:12.628523 & 0.190 & 0.409 & 2.509$\pm$0.044 & 0.060$\pm$0.200 & 6.950$\pm$0.160 & 13 \\
    MT Ori & 2015.180000 & 05:35:17.9522494 & $-$05:22:45.435871 & 0.384 & 0.828 & 2.646$\pm$0.041 & 3.820$\pm$0.100 & 1.600$\pm$0.170 & 13 \\
    V1046 Ori & 2015.200000 & 05:35:21.8682172 & $-$04:29:39.000548 & 0.408 & 0.887 & 2.643$\pm$0.075 & 1.880$\pm$0.090 & 1.200$\pm$0.140 & 13 \\
    HD 37150 & 2015.190000 & 05:36:15.0289993 & $-$05:38:52.516895 & 0.399 & 0.862 & 2.536$\pm$0.046 & 1.320$\pm$0.050 & -0.560$\pm$0.120 & 13 \\
    TYC 5346-538-1 & 2015.200000 & 05:42:33.6768420 & $-$08:07:15.182416 & 0.217 & 0.426 & 2.348$\pm$0.069 & 0.680$\pm$0.090 & -0.510$\pm$0.250 & 13 \\
    HD 290862 & 2015.210000 & 05:46:43.3856400 & $+$00:04:36.034276 & 0.790 & 1.132 & 2.197$\pm$0.545 & 0.350$\pm$0.270 & 0.830$\pm$0.830 & 13 \\
    {[SSC75] M 78 11} & 2015.210000 & 05:46:45.3379210 & $+$00:02:40.293196 & 0.342 & 0.574 & 2.547$\pm$0.034 & 0.010$\pm$0.100 & -0.490$\pm$0.080 & 13 \\
    V1355 Ori & 2023.250000 & 06:02:40.3788999 & $-$00:51:36.953609 & 0.161 & 0.374 & 7.955$\pm$0.031 & 12.514$\pm$0.028 & 11.400$\pm$0.061 & 2 \\
    VY CMa & 2006.530000 & 07:22:58.3259671 & $-$25:46:03.062417 & 10.150 & 10.336 & 0.830$\pm$0.080 & -2.210$\pm$0.060 & 2.290$\pm$0.300 & 14 \\
    XY UMa & 2023.250000 & 09:09:55.8026472 & $+$54:29:13.477129 & 0.399 & 0.662 & 14.729$\pm$0.571 & -49.789$\pm$0.427 & -182.889$\pm$0.328 & 2 \\
    FF UMa & 2023.250000 & 09:33:46.4791651 & $+$62:49:39.738521 & 0.128 & 0.130 & 8.470$\pm$0.098 & -20.057$\pm$0.073 & -22.157$\pm$0.072 & 2 \\
    DM UMa & 2023.250000 & 10:55:43.4245688 & $+$60:28:09.544893 & 0.184 & 0.213 & 5.473$\pm$0.075 & -37.848$\pm$0.050 & -7.598$\pm$0.054 & 2 \\
    S Crt & 2005.798800 & 11:52:44.9697296 & $-$07:35:48.086977 & 1.833 & 13.336 & 2.330$\pm$0.130 & -3.170$\pm$0.220 & -5.410$\pm$0.220 & 15 \\
    RS CVn & 2023.250000 & 13:10:36.8122478 & $+$35:56:06.066804 & 0.097 & 0.181 & 7.488$\pm$0.104 & -49.973$\pm$0.051 & 20.643$\pm$0.107 & 2 \\
    BH CVn & 1993.108800 & 13:34:47.7594787 & $+$37:10:56.760002 & 0.373 & 0.455 & 22.210$\pm$0.450 & 85.496$\pm$0.131 & -9.220$\pm$0.160 & 4 \\
    S CrB & 2000.000000 & 15:21:23.9560824 & $+$31:22:02.573040 & - & - & 2.360$\pm$0.230 & -9.060$\pm$0.230 & -12.520$\pm$0.290 & 16 \\
    RS UMi & 2023.250000 & 15:50:49.4500275 & $+$72:12:40.428941 & 0.127 & 0.327 & 2.187$\pm$0.014 & 3.340$\pm$0.017 & -8.684$\pm$0.180 & 2 \\
    sig CrB & 1990.001400 & 16:14:41.0675052 & $+$33:51:31.873993 & 0.104 & 0.124 & 43.930$\pm$0.100 & -267.048$\pm$0.037 & -86.660$\pm$0.050 & 4 \\
    U Her & 2000.000000 & 16:25:47.4716736 & $+$18:53:32.855640 & - & - & 3.740$\pm$0.610 & -14.980$\pm$0.290 & -9.230$\pm$0.320 & 16 \\
    Haro 1-6 & 2007.990000 & 16:26:03.0160198 & $-$24:23:36.405235 & 0.596 & 0.856 & 7.385$\pm$0.234 & -19.630$\pm$0.190 & -26.920$\pm$0.130 & 17 \\
    DoAr 51 & 2014.750000 & 16:32:11.7909539 & $-$24:40:21.962916 & 0.534 & 0.836 & 6.983$\pm$0.050 & -4.800$\pm$0.080 & -23.110$\pm$0.110 & 17 \\
    W 40 IRS 5 & 2014.950000 & 18:31:14.8227619 & $-$02:03:50.157308 & 0.156 & 0.291 & 2.302$\pm$0.063 & 0.186$\pm$0.053 & -6.726$\pm$0.121 & 18 \\
    HD 179094 & 2021.125743 & 19:08:25.5551833 & $+$52:25:31.480133 & 1.461 & 0.998 & 14.300$\pm$0.338 & -102.158$\pm$2.449 & -55.217$\pm$0.570 & 5 \\
    RR Aql & 2000.000000 & 19:57:36.0597288 & $-$01:53:11.338656 & - & - & 1.580$\pm$0.400 & -25.110$\pm$0.740 & -49.820$\pm$0.540 & 16 \\
    Cyg X-1 & 1991.249800 & 19:58:21.6784714 & $+$35:12:05.840028 & 0.308 & 0.368 & 0.730$\pm$0.300 & -3.787$\pm$0.172 & -6.250$\pm$0.210 & 4 \\
    Cyg X-1 & 2009.570000 & 19:58:21.6727548 & $+$35:12:05.725022 & 0.500 & 0.500 & 0.547$\pm$0.041 & -3.700$\pm$0.080 & -6.420$\pm$0.140 & 19 \\
    HD 199178 & 1993.793300 & 20:53:53.6368109 & $+$44:23:11.085234 & 0.332 & 0.397 & 8.590$\pm$0.330 & 26.595$\pm$0.407 & -1.240$\pm$0.430 & 4 \\
    HD 199178 & 2020.895230 & 20:53:53.7035929 & $+$44:23:11.062287 & 0.361 & 0.376 & 8.953$\pm$0.093 & 26.381$\pm$0.225 & -0.876$\pm$0.177 & 20 \\
    SS Cyg & 2011.566100 & 21:42:42.9230791 & $+$43:35:10.253553 & 0.216 & 0.215 & 8.800$\pm$0.120 & 112.420$\pm$0.070 & 33.380$\pm$0.070 & 21 \\
    FF Aqr & 2023.250000 & 22:00:36.4739751 & $-$02:44:27.109979 & 0.126 & 0.212 & 4.564$\pm$0.072 & 32.774$\pm$0.043 & -11.826$\pm$0.061 & 2 \\
    AR Lac & 1992.435300 & 22:08:40.8558984 & $+$45:44:31.751477 & 0.274 & 0.361 & 23.970$\pm$0.370 & -52.080$\pm$0.126 & 47.030$\pm$0.190 & 4 \\
    AR Lac & 2020.895340 & 22:08:40.7137036 & $+$45:44:33.089029 & 0.161 & 0.126 & 23.529$\pm$0.130 & -51.884$\pm$0.265 & 46.830$\pm$0.224 & 20 \\
    IM Peg & 1992.917200 & 22:53:02.2762503 & $+$16:50:28.492233 & 0.394 & 0.428 & 10.280$\pm$0.620 & -20.587$\pm$0.459 & -27.530$\pm$0.400 & 4 \\
    IM Peg & 2005.086900 & 22:53:02.2586161 & $+$16:50:28.159982 & 0.434 & 0.426 & 10.370$\pm$0.074 & -20.833$\pm$0.090 & -27.267$\pm$0.095 & 22 \\
    SZ Psc & 2021.125933 & 23:13:23.8098299 & $+$02:40:32.164513 & 0.650 & 0.935 & 10.134$\pm$0.243 & 13.158$\pm$0.843 & 26.790$\pm$0.235 & 5 \\
    PZ Cas & 2006.299800 & 23:44:03.2815716 & $+$61:47:22.187580 & 3.013 & 3.086 & 0.356$\pm$0.026 & -3.700$\pm$0.200 & -2.000$\pm$0.300 & 23 \\
\end{longtable}
\begin{tablenotes}
    \makebox[1.25\textwidth]{
        \footnotesize
        \begin{minipage}{1.25\textwidth}
            \item $\sigma_{\alpha*}$ and $\sigma_{\delta}$ denote position errors in RA$\cdot\cos$(DEC) and DEC directions, respectively.
            Positional data without uncertainties (``-'') were not used in the solutions.
            \item References: (1) \citet{2011PASJ...63...63N}; (2) Zhang et al. in prep.; (3) \citet{2010ApJ...721..267A}; (4) \citet{1999A&A...344.1014L}; (5) Jiang et al. in prep.; (6) \citet{2011ApJ...737..104P}; (7) \citet{2014Sci...345.1029M}; (8) \citet{2012ApJ...747...18T}; (9) \citet{2018ApJ...859...33G}; (10) \citet{2007ApJ...671.1813T}; (11) \citet{2007ApJ...671..546L}; (12) \citet{2014PASJ...66..101N}; (13) \citet{2017ApJ...834..142K}; (14) \citet{2012ApJ...744...23Z}; (15) \citet{2008PASJ...60.1013N}; (16) \citet{2007A&A...472..547V}; (17) \citet{2017ApJ...834..141O}; (18) \citet{2017ApJ...834..143O}; (19) \citet{2011ApJ...742...83R}; (20) \citet{2023MNRAS.524.5357C}; (21) \citet{2013Sci...340..950M}; (22) \citet{2015CQGra..32v4021B}; (23) \citet{2013ApJ...774..107K}.
        \end{minipage}
    }
\end{tablenotes}
\end{ThreePartTable}

\end{landscape}

\begin{table}[h!]
    \caption{\label{table:data2}One-epoch VLBI data used for CRF alignment}
    \centering
    \begin{tabular}{lcrrrrr}
        \hline\hline
        \makebox[62pt][c]{Star} & Epoch & \makebox[73pt][c]{RA} & \makebox[72pt][c]{DEC} & \makebox[26pt][c]{$\sigma_{\alpha*}$} & \makebox[26pt][c]{$\sigma_{\delta}$} & Ref. \\
        & & \makebox[73pt][c]{(h:m:s)} & \makebox[72pt][c]{(d:m:s)} & \makebox[26pt][c]{(mas)} & \makebox[26pt][c]{(mas)} & \\
        \hline
        UV Psc & 2020.015060 & 01:16:55.2328601 & $+$06:48:42.545459 & 0.286 & 0.700 & 1 \\
        HD 8357 & 2020.015070 & 01:22:56.8823023 & $+$07:25:13.949756 & 0.283 & 0.694 & 1 \\
        LS I +61 303 & 2020.015200 & 02:40:31.6633188 & $+$61:13:45.588648 & 0.315 & 0.701 & 1 \\
        RZ Cas & 2020.015200 & 02:48:55.5183194 & $+$69:38:04.194762 & 0.315 & 0.714 & 1 \\
        UX Ari & 2020.015420 & 03:26:35.4456180 & $+$28:42:52.171571 & 0.429 & 0.825 & 1 \\
        UX Ari & 2020.015420 & 03:26:35.4455928 & $+$28:42:52.170397 & 0.297 & 0.700 & 1 \\
        UX Ari & 2020.015420 & 03:26:35.4455030 & $+$28:42:52.171276 & 0.431 & 0.830 & 1 \\
        UX Ari & 2020.015420 & 03:26:35.4454778 & $+$28:42:52.170160 & 0.299 & 0.705 & 1 \\
        HD 22468 & 2020.015450 & 03:36:47.2457954 & $+$00:35:12.688282 & 0.282 & 0.692 & 1 \\
        HD 283447 & 2020.015440 & 04:14:12.9344146 & $+$28:12:11.807694 & 0.376 & 0.829 & 1 \\
        HD 283447 & 2020.015440 & 04:14:12.9344304 & $+$28:12:11.809246 & 0.291 & 0.713 & 1 \\
        HD 283447 & 2020.015440 & 04:14:12.9346284 & $+$28:12:11.809616 & 0.375 & 0.828 & 1 \\
        HD 283447 & 2020.015440 & 04:14:12.9346440 & $+$28:12:11.811280 & 0.291 & 0.713 & 1 \\
        B Per & 2020.015440 & 04:18:14.7095050 & $+$50:17:42.686621 & 0.284 & 0.695 & 1 \\
        V410 Tau & 2020.015460 & 04:18:31.1212018 & $+$28:27:15.653570 & 0.287 & 0.703 & 1 \\
        HD 283572 & 2020.015470 & 04:21:58.8615370 & $+$28:18:05.984222 & 0.284 & 0.695 & 1 \\
        T Tau & 2020.015480 & 04:21:59.4376870 & $+$19:32:05.551793 & 0.306 & 0.739 & 1 \\
        CoKu HP Tau G2 & 2020.015490 & 04:35:54.1727731 & $+$22:54:13.186156 & 0.303 & 0.724 & 1 \\
        CoKu HP Tau G2 & 2020.015490 & 04:35:54.1728058 & $+$22:54:13.186037 & 0.304 & 0.722 & 1 \\
        Brun 334 & 2020.015600 & 05:34:39.7587494 & $-$05:24:25.609900 & 0.307 & 0.739 & 1 \\
        TYC 5346-538-1 & 2020.015610 & 05:42:33.6769745 & $-$08:07:15.184528 & 0.337 & 0.788 & 1 \\
        SV Cam & 2020.014610 & 06:41:19.5133901 & $+$82:15:59.418299 & 0.283 & 0.695 & 1 \\
        AR Mon & 2021.678721 & 07:20:48.4647784 & $-$05:15:35.943515 & 0.109 & 0.253 & 2 \\
        AR Mon & 2021.804318 & 07:20:48.4648528 & $-$05:15:35.945308 & 0.075 & 0.164 & 2 \\
        54 Cam & 2020.013580 & 08:02:35.6931432 & $+$57:16:23.871857 & 0.300 & 0.700 & 1 \\
        IL Hya & 2020.013480 & 09:24:48.9564576 & $-$23:49:35.361127 & 0.302 & 0.740 & 1 \\
        HU Vir & 2020.013710 & 12:13:20.6754857 & $-$09:04:46.858105 & 0.306 & 0.740 & 1 \\
        DK Dra & 2020.013620 & 12:15:41.4523546 & $+$72:33:03.809124 & 0.308 & 0.702 & 1 \\
        RS CVn & 2020.013700 & 13:10:36.8261232 & $+$35:56:05.997422 & 0.283 & 0.694 & 1 \\
        BH CVn & 2020.013700 & 13:34:47.9537587 & $+$37:10:56.490899 & 0.319 & 0.726 & 1 \\
        BH CVn & 2020.013700 & 13:34:47.9537057 & $+$37:10:56.489448 & 0.318 & 0.737 & 1 \\
        del Lib & 2020.014050 & 15:00:58.2632962 & $-$08:31:08.335088 & 0.312 & 0.702 & 1 \\
        del Lib & 2020.014050 & 15:00:58.2632419 & $-$08:31:08.333656 & 0.350 & 0.827 & 1 \\
        sig CrB & 2020.014160 & 16:14:40.4246654 & $+$33:51:29.234981 & 0.282 & 0.691 & 1 \\
        Haro 1-6 & 2020.014290 & 16:26:02.9989615 & $-$24:23:36.731260 & 0.319 & 0.789 & 1 \\
        DoAr 51 & 2020.014290 & 16:32:11.7895704 & $-$24:40:22.082106 & 0.322 & 0.803 & 1 \\
        DoAr 51 & 2020.014290 & 16:32:11.7892778 & $-$24:40:22.087733 & 0.324 & 0.808 & 1 \\
        HD 199178 & 2020.014520 & 20:53:53.7011153 & $+$44:23:11.055530 & 0.394 & 0.753 & 1 \\
        SS Cyg & 2020.014510 & 21:42:43.0100261 & $+$43:35:10.528408 & 0.305 & 0.714 & 1 \\
        AR Lac & 2020.014600 & 22:08:40.7168030 & $+$45:44:33.030622 & 0.297 & 0.701 & 1 \\
        AR Lac & 2020.014600 & 22:08:40.7167838 & $+$45:44:33.030877 & 0.282 & 0.692 & 1 \\
        IM Peg & 2020.014690 & 22:53:02.2364210 & $+$16:50:27.746012 & 0.282 & 0.692 & 1 \\
        SZ Psc & 2020.014750 & 23:13:23.8079419 & $+$02:40:32.129994 & 0.471 & 1.106 & 1 \\
        SZ Psc & 2020.014750 & 23:13:23.8079614 & $+$02:40:32.130487 & 0.283 & 0.692 & 1 \\
        \hline
    \end{tabular}
    \tablefoot{References: (1) \citet{2023A&A...676A..11L}; (2) Zhang et al., in prep.
    }
\end{table}

\FloatBarrier
\clearpage

\section{\texorpdfstring{Iteration of solution $0<G<10.5$}{Iteration of solution 0<G<10.5}}
\label{app:iter_10.5}

\begin{figure}[ht]
    \centering
    \includegraphics[width=0.97\hsize]{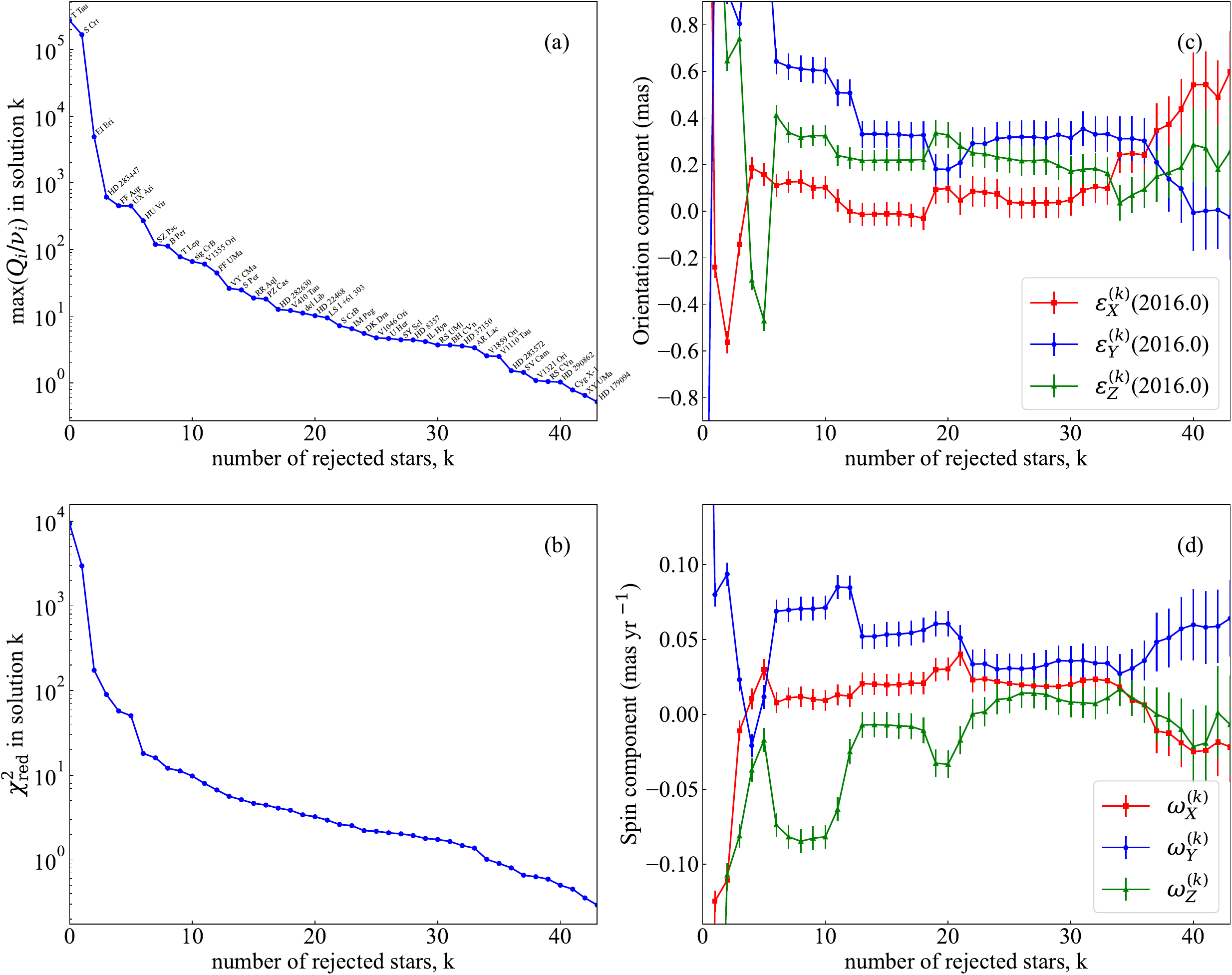}
    \caption{The evolution of loss and CRF link parameters with rejected star number $k$ grows on the dataset $0<G<10.5$.}
    \label{fig:evo_bright}
\end{figure}

\clearpage

\section{\texorpdfstring{Statistics of solution $0<G<10.5$}{Statistics of solution 0<G<10.5}}
\label{app:stat_10.5}

\begin{table}[ht]
    \caption{\label{table:stat}Statistics of solution $0<G<10.5$ for each star}
    \centering
    \setlength{\tabcolsep}{12pt}
    \begin{tabular}{lrrrr}
        \hline\hline
        \makebox[62pt][c]{Star} & \makebox[10pt][c]{$\nu$} & \makebox[25pt][c]{$Q_i/\nu_i$} & \makebox[30pt][c]{$E_i$} & \makebox[30pt][c]{$\Omega_i$} \\
        & & & \makebox[30pt][c]{(mas$^{-2}$)} & \makebox[30pt][c]{(mas$^{-2}$yr$^2$)} \\
        \hline
        SY Scl          &     5 &   4.65 &    0.0 &   93.8 \\
        S Per           &     5 &  24.82 &    0.0 &   43.7 \\
        LS I +61 303    &     7 &  11.78 &   22.2 & 5082.8 \\
        HD 22468        &     7 &  11.59 &   20.3 & 1435.6 \\
        V410 Tau        &     7 &  12.05 &  412.4 & 2725.2 \\
        HD 283572       &    12 &   1.54 &   73.8 & 4719.7 \\
        V1110 Tau       &     5 &   3.36 &   28.3 &  446.5 \\
        HD 282630       &     5 &  12.62 &   47.2 &  147.0 \\
        V1321 Ori       &     5 &   0.70 &   33.6 &   86.1 \\
        V1046 Ori       &     5 &   4.53 &    7.2 &  136.4 \\
        HD 37150        &     5 &   3.46 &    7.5 &  261.1 \\
        HD 290862       &     5 &   0.97 &    2.4 &   16.2 \\
        VY CMa          &     5 &  26.17 &    0.0 &   14.9 \\
        BH CVn          &     9 &   3.49 &   27.5 &  868.6 \\
        S CrB           &     3 &   7.19 &    0.0 &   27.4 \\
        U Her           &     3 &   4.70 &    0.0 &   20.6 \\
        RR Aql          &     3 &  18.76 &    0.0 &    5.1 \\
        Cyg X-1         &    10 &   1.60 &   13.4 & 4747.5 \\
        HD 199178       &    12 &   0.96 &   36.0 & 4232.6 \\
        AR Lac          &    14 &   3.08 &  153.7 & 3794.5 \\
        IM Peg          &    12 &   6.25 &   23.2 &  505.1 \\
        PZ Cas          &     5 &  18.14 &    0.2 &   53.3 \\
        UV Psc          &     2 &   0.17 &   12.9 &  207.4 \\
        HD 8357         &     7 &   5.14 &   13.3 &  460.0 \\
        RZ Cas          &     7 &   1.45 &   39.9 &  968.3 \\
        SV Cam          &     2 &   0.60 &   13.8 &  222.2 \\
        54 Cam          &     2 &   0.40 &   12.3 &  197.4 \\
        IL Hya          &     2 &   4.24 &   11.4 &  183.0 \\
        DK Dra          &     2 &   6.79 &   11.3 &  181.9 \\
        RS CVn          &     7 &   0.47 &   99.4 & 5050.1 \\
        del Lib         &     4 &  11.06 &    1.0 &   16.3 \\
        V1859 Ori       &     5 &   3.47 &   32.2 & 2591.0 \\
        AR Mon          &     4 &   0.47 &   57.2 & 1898.8 \\
        XY UMa          &     5 &   0.59 &   41.3 & 1113.0 \\
        DM UMa          &     5 &   0.93 &   40.6 & 3097.9 \\
        RS UMi          &     5 &   3.74 &   49.4 & 5551.0 \\
        HD 179094       &     5 &   0.51 &    1.3 &   36.1 \\
        \hline
    \end{tabular}
    \tablefoot{$\nu$ is the total VLBI data dimension of each star, e.g., 5 for five-parameter measurement and 2 for one-epoch measurement.
    $Q_i/\nu_i$ is the normalized loss.
    $E_i$ and $\Omega_i$ are weights (amounts of information contributed) in orientation and spin respectively.
    Only accepted stars are listed in this table.
    }
\end{table}

\clearpage

\section{Radio star sky distribution}
\label{app:distri}

\begin{figure}[ht]
    \centering
    \includegraphics[width=\hsize]{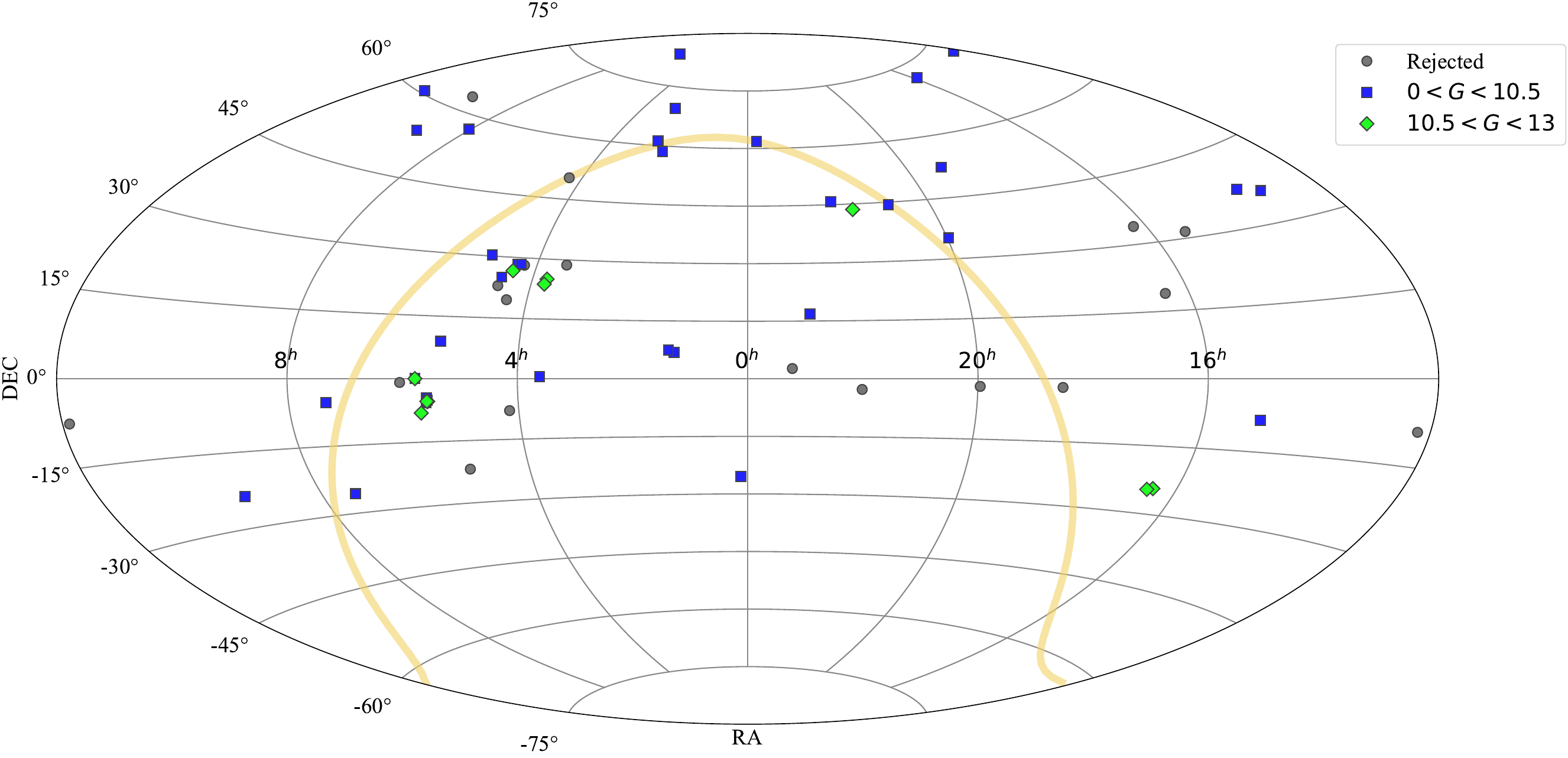}
    \caption{The sky distribution (Aitoff projection in equatorial coordinates) of the radio stars used in this study.
    The blue squares and green diamonds denote samples used in solution $0<G<10.5$ and $10.5<G<13$, respectively.
    The rejected stars are plotted as grey dots.
    The Galactic Equator is plotted in yellow.}
    \label{fig:star_distri}
\end{figure}

The sky distribution of the radio stars is shown in Fig. \ref{fig:star_distri}.
The whole sample lacks stars in the southern sky ($<-30^{\circ}$).
The number of stars between $G=9$ and 13 is small and the stars are unevenly distributed, especially concentrated near the galactic plane, which may cause selection biases.
A larger number of more evenly distributed available radio stars will help improve our results and reduce the impact of potential selection biases.

\twocolumn

\section{Mathematics for orientation-only solution}
\label{app:math}

The solution is based on Sect. 2 of \citet{2024MNRAS.529.2062Z}, and the modifications needed for the orientation-only solution are as follows:

Proper motions of Gaia data $\boldsymbol{\mu}=[\mu_{\alpha*}, \mu_{\delta}]^{\prime}$ need to be corrected with the a priori spin parameter $\boldsymbol{\omega}=[\omega_{X}, \omega_{Y}, \omega_{Z}]^{\prime}$ before the solution
\begin{equation}
    \boldsymbol{\mu}_{\rm corrected}=\boldsymbol{\mu}-\boldsymbol{L}\boldsymbol{\omega} \ ,
\end{equation}
where
\begin{equation}
    \boldsymbol{L}=\begin{bmatrix}
        \mathrm{c} \alpha_{i} \mathrm{s} \delta_{i} & \mathrm{s} \alpha_{i} \mathrm{s} \delta_{i} & -\mathrm{c} \delta_{i} \\
        -\mathrm{s} \alpha_{i} & \mathrm{c} \alpha_{i} & 0
        \end{bmatrix} \ ;
\end{equation}

The parameters to be estimated is changed to $\boldsymbol{x}=[\varepsilon_{X}(T), \varepsilon_{Y}(T), \varepsilon_{Z}(T)]^{\prime}$;

Because of the shape of $\boldsymbol{x}$ is changed from $6\times1$ to $3\times1$, the coefficient matrix $\boldsymbol{K}_{i}$ also needs to be changed to
\begin{equation}
    \boldsymbol{K}_{i}=\begin{bmatrix}
        \mathrm{c} \alpha_{i} \mathrm{s} \delta_{i} & \mathrm{s} \alpha_{i} \mathrm{s} \delta_{i} & -\mathrm{c} \delta_{i} \\
        -\mathrm{s} \alpha_{i} & \mathrm{c} \alpha_{i} & 0\\
        0 & 0 & 0 \\
        0 & 0 & 0 \\
        0 & 0 & 0
        \end{bmatrix}
\end{equation}
to match the shape of $\boldsymbol{x}$.

\section{Solution correlation coefficient matrices}
\label{app:corr}

\subsubsection*{\texorpdfstring{Solution $0<G<13$}{Solution 0<G<13}}
\begin{multline}
    \mathrm{corr}[\varepsilon_{X}(T), \varepsilon_{Y}(T), \varepsilon_{Z}(T), \omega_{X}, \omega_{Y}, \omega_{Z}]=\\\begin{bmatrix}
        +1.000 & +0.191 & +0.204 & -0.102 & -0.120 & -0.206 \\
        \cdots & +1.000 & +0.277 & -0.139 & -0.214 & -0.105 \\
        \cdots & \cdots & +1.000 & -0.177 & -0.090 & -0.154 \\
        \cdots & \cdots & \cdots & +1.000 & -0.032 & +0.099 \\
        \cdots & \cdots & \cdots & \cdots & +1.000 & -0.021 \\
        \cdots & \cdots & \cdots & \cdots & \cdots & +1.000
    \end{bmatrix}
\end{multline}

\subsubsection*{\texorpdfstring{Solution $0<G<10.5$}{Solution 0<G<10.5}}
\begin{multline}
    \mathrm{corr}[\varepsilon_{X}(T), \varepsilon_{Y}(T), \varepsilon_{Z}(T), \omega_{X}, \omega_{Y}, \omega_{Z}]=\\\begin{bmatrix}
        +1.000 & +0.134 & +0.192 & -0.166 & -0.100 & -0.237 \\
        \cdots & +1.000 & +0.255 & -0.132 & -0.279 & -0.091 \\
        \cdots & \cdots & +1.000 & -0.213 & -0.068 & -0.280 \\
        \cdots & \cdots & \cdots & +1.000 & -0.043 & +0.120 \\
        \cdots & \cdots & \cdots & \cdots & +1.000 & -0.043 \\
        \cdots & \cdots & \cdots & \cdots & \cdots & +1.000
    \end{bmatrix}
\end{multline}

\subsubsection*{\texorpdfstring{Orientation-only solution $0<G<13$}{Orientation-only solution 0<G<13}}
\begin{equation}
    \mathrm{corr}[\varepsilon_{X}(T), \varepsilon_{Y}(T), \varepsilon_{Z}(T)]=\begin{bmatrix}
        +1.000 & +0.142 & +0.158 \\
        \cdots & +1.000 & +0.239 \\
        \cdots & \cdots & +1.000
    \end{bmatrix}
\end{equation}

\subsubsection*{\texorpdfstring{Orientation-only solution $0<G<10.5$}{Orientation-only solution 0<G<10.5}}
\begin{equation}
    \mathrm{corr}[\varepsilon_{X}(T), \varepsilon_{Y}(T), \varepsilon_{Z}(T)]=\begin{bmatrix}
        +1.000 & +0.069 & +0.102 \\
        \cdots & +1.000 & +0.206 \\
        \cdots & \cdots & +1.000
    \end{bmatrix}
\end{equation}

\subsubsection*{\texorpdfstring{Orientation-only solution $10.5<G<13$}{Orientation-only solution 10.5<G<13}}
\begin{equation}
    \mathrm{corr}[\varepsilon_{X}(T), \varepsilon_{Y}(T), \varepsilon_{Z}(T)]=\begin{bmatrix}
        +1.000 & +0.414 & +0.345 \\
        \cdots & +1.000 & +0.397 \\
        \cdots & \cdots & +1.000
    \end{bmatrix}
\end{equation}

\end{appendix}
\end{document}